\documentclass[12pt]{article}

\usepackage[top=60pt,bottom=70pt,left=75pt,right=75pt]{geometry}

\usepackage{amsmath, amssymb, amsfonts, latexsym, euscript, mathalfa, mathrsfs}
\usepackage{graphicx, color, epsfig, xcolor }

\usepackage{setspace, sectsty}

\usepackage{braket, cancel}

\usepackage[debug,pageanchor=false]{hyperref}
\definecolor{link}{rgb}{.8,.15,.1}
\hypersetup{colorlinks=true,linkcolor=link,citecolor=link,urlcolor=link,linktocpage}

\renewcommand{\arraystretch}{1.2}

\makeatletter
\@addtoreset{equation}{section}
\makeatother

\def\be{\begin{equation}}
\def\ee{\end{equation}}

\newcommand{\p}{\dot{\alpha}^2 - 2 \alpha \ddot{\alpha}}

\begin{document}

\begin{titlepage}

\begin{flushright} \small
UUITP-26/18
\end{flushright}

\begin{center}

%\vskip .5in %.3in
\noindent

{\Large \bf AdS$_2 \times S^7$ solutions from D0 -- F1 -- D8 intersections}

\bigskip\medskip

Giuseppe Dibitetto and Achilleas Passias\\

\bigskip\medskip
{\small

Department of Physics and Astronomy, Uppsala University,\\
Box 516, SE-75120 Uppsala, Sweden

}

\vskip .5cm %.3cm
{\small \tt giuseppe.dibitetto@physics.uu.se, achilleas.passias@physics.uu.se }

\vskip .9cm %.6cm
     	{\bf Abstract }
\vskip .1in
\end{center}

\noindent We study an exhaustive analytic class of massive type IIA backgrounds preserving sixteen real supercharges and enjoying $\textrm{SL}(2,\mathbb{R})\times\textrm{SO}(8)$ 
 bosonic symmetry. The corresponding geometry is described by $\textrm{AdS}_{2}\times S^{7}$ warped over a line, which turns out to emerge from taking the 
 near-horizon limit of D0 -- F1 -- D8 intersections. 
 By studying the singularity structure of these solutions we find the possible presence of localized O8/D8 sources, as well as of fundamental strings smeared over the $S^{7}$.
 Finally we discuss the relation between the aforementioned solutions and the known $\textrm{AdS}_{7}\times S^{2}$ class through double analytic continuation.

\vfill
\eject

\end{titlepage}

\tableofcontents

\section{Introduction}

Ever since the discovery of the AdS/CFT correspondence \cite{Maldacena:1997re}, there has been a lot of effort devoted to the classification of supersymmetric AdS vacua 
in string theory. While there exist very few examples enjoying maximal supersymmetry, a much richer structure opens up once we look into backgrounds preserving
half-maximal supersymmetry, \emph{i.e.} sixteen real supercharges. This fact is mainly due to the possibility of having solutions of the form 
$\textrm{AdS}_{d+1}\times\mathcal{M}_{9-d}$, where the corresponding geometries include a non-trivial warping.

When focusing on two-dimensional holography though, a possible $\textrm{AdS}_{2}/\textrm{CFT}_{1}$ correspondence is often thought of as far less 
understood than its higher-dimensional counterparts. In particular, many issues and exotic features of gravity in a nearly-$\textrm{AdS}_{2}$ geometry
are encountered along the way \cite{Strominger:1998yg,Maldacena:1998uz}. Among these we mention the presence of multiple disconnected time-like boundary components, which may represent a crucial 
obstruction to identifying a correct holographic dictionary in the first place.
A renovated interest in the topic has been sparked by the so-called SYK model \cite{Sachdev:1992fk} and its possible realizations in high-energy theory 
(see \cite{Maldacena:2016hyu} and references therein), along with the novel issues which were recently discussed in \cite{Harlow:2018tqv} within this context.

The general challenge posed by two-dimensional holography may be viewed as a motivation for looking into supersymmetric $\textrm{AdS}_{2}$ vacua in 
string theory with a clear brane picture, as those might shed a light on the unresolved issues concerning the $\textrm{AdS}_{2}/\textrm{CFT}_{1}$ correspondence
within a controlled framework.
The aim of the present work is precisely that of enriching the landscape of supersymmetric $\textrm{AdS}_{2}$ string vacua by presenting a new class
of such solutions in massive type IIA string theory.

The class under consideration here will be identified with geometries given by warped products of $\textrm{AdS}_{2}$ and an 8-manifold constructed as
a round seven-sphere fibered over a line. The existence of such solutions may be inferred from ``double analytic continuation'' arguments which would 
relate them to the ones in the $\textrm{AdS}_{7}\times S^{2}$ class of \cite{Apruzzi:2013yva,Apruzzi:2015wna}, in the same way as type IIB geometries
of the form $\textrm{AdS}_{2}\times S^{6}$ warped over a Riemann surface were argued in \cite{Corbino:2017tfl} to be related to previously known backgrounds
$\textrm{AdS}_{6}\times S^{2}$ warped over a Riemann surface, through the aforementioned double analytic continuation.

The paper is organized as follows. We start out by reviewing some facts concerning D0 -- F1 -- D8 intersections in massive type IIA string theory and 
relate them to $\frac14$-BPS supergravity solutions discussed in \cite{Imamura:2001cr}. Subsequently, we show how to make an educated guess in the general
\emph{Ansatz} which directly produces solutions with enhanced supersymmetry thus realizing $\textrm{AdS}_{2}$ geometry.
After integrating the obtained differential equations, we discuss the relation of the obtained solutions to the aforementioned $\textrm{AdS}_{7}$ counterparts through
double analytic continuation. Then, we discuss the possible singularity structures as well as the range of the warp coordinate.
We conclude with some further speculations concerning the possible holographic interpretation of our work.

\section{AdS$_2$ solutions from D0 -- F1 -- D8 intersections}

D0 -- D8 bound states were considered in \cite{Bachas:1997kn,Danielsson:1997gi,Witten:2000mf, Fujii:2001wp} as UV descriptions of $\mathcal{N}=(8,0)$ superconformal quantum mechanics.
Such bound states require a non-trivial $B$-field sourced by a fundamental string.
Note that, contrary to all other D-branes, strings cannot end on a D-particle due to charge conservation issues \cite{Strominger:1995ac,Townsend:1996em}. 
However the situation changes in presence of D8-branes in the background, where in fact an F1 stretched between a D$0$ and the D$8$ has to be formed whenever
the D$0$ crosses a D$8$ \cite{Danielsson:1997wq}. This physical process can be understood as a dual version of the Hanany-Witten (HW) effect \cite{Hanany:1996ie}.

D0 -- F1 -- D8 brane systems were exhaustively studied in \cite{Imamura:2001cr, Massar:1999sb} as a class of $\frac{1}{4}$-BPS solutions in massive type IIA supergravity.
The explicit set-up is summarized in table~\ref{Table:D0F1D8}.
\begin{table}[h!]
\renewcommand{\arraystretch}{1}
\begin{center}
\scalebox{1}[1]{
\begin{tabular}{c||c||c | c c c c c c c c}
%\hline
branes & $t$ & $y$ & $r$ & $\theta^{1}$ & $\theta^{2}$ & $\theta^{3}$ & $\theta^{4}$ & $\theta^{5}$ & $\theta^{6}$ & $\theta^{7}$ \\
\hline \hline
D0 & $\times$ & $-$ & $-$ & $-$ & $-$ & $-$ & $-$ & $-$ & $-$ & $-$ \\
%\hline
F1 & $\times$ & $\times$ & $-$ & $-$ & $-$ & $-$ & $-$ & $-$ & $-$ & $-$ \\
%\hline
D8 & $\times$ & $-$ & $\times$ & $\times$ & $\times$ & $\times$ & $\times$ & $\times$ & $\times$ & $\times$ \\
\end{tabular}
}
\end{center}
\caption{{\it The brane picture underlying the $\mathcal{N}=(8,0)$ superconformal quantum mechanics described by a D0 -- F1 -- D8 system. The above system is $\frac{1}{4}$-BPS.
We anticipate that the radial coordinate realizing the dual $\textrm{AdS}_{2}$ geometry turns out to be a combination of $y$ \& $r$.}} \label{Table:D0F1D8}
\end{table}

The complete \emph{Ansatz} for the 10D fields of massive type IIA supergravity can be completely specified in terms of two arbitrary functions of the $(y,r)$
coordinates, denoted by $S$ \& $K$. This reads
\begin{equation}
\label{Imamura_Ansatz}
\begin{array}{lclc}
ds^2_{10} & = &  -S^{-1} K^{-1/2} \, dt^2 + S^{-1}K^{1/2} \, dy^2 + K^{\frac12}\,(dr^2 + r^2 d\Omega_{(7)}^2) & , \\[1mm]
e^\Phi & = & g_s \, K^{3/4}S^{-1/2} & , \\[1mm]
F_{(0)} & = & m & , \\[1mm]
C_{(1)} & = & g^{-1}_s \, K^{-1} \,dt & , \\[1mm]
B_{(2)} & = & S^{-1} \, dt \wedge dy & , 
\end{array}
\end{equation}
where $g_{s}>0$ is the string coupling and $d\Omega_{(7)}^2$ denotes the metric of a unit-radius seven-sphere.

For non-zero Romans' mass $m$ the full set of 10D equations of motion and Bianchi identities is implied by the following non-linear PDE's
\begin{subequations}
\begin{align}
m g_s \,K -  \frac{\partial S}{\partial y}&= 0 \ , \label{Massive_Keqn}\\
\Delta_{(8)} S + \frac{1}{2} \frac{\partial^2}{\partial y^2} S^2 &= 0 \label{Massive_Seqn} \ ,
\end{align}
\end{subequations}
where $\Delta_{(8)}$ is the Laplace operator on $\mathbb{R}^{8}$, \emph{i.e.} the common transeverse space to the D-particle and the string.
Its rotationally invariant form expressed in polar coordinates reads
\begin{equation}
\Delta_{(8)} \equiv \frac{1}{r^7} \frac{\partial}{\partial r} r^7 \frac{\partial}{\partial r} \ .
\end{equation} 

Note that in the massless limit $K$ is no longer determined by the equation \eqref{Massive_Keqn}, but it must instead satisfy the following PDE
\begin{equation}
\label{Massless_Keqn}
\Delta_{(8)} K + S \frac{\partial^2}{\partial y^2} K = 0 \ .
\end{equation}
A particular solution to \eqref{Massive_Seqn} \& \eqref{Massless_Keqn} is given by the semilocalized D0 -- F1 intersection constructed by following the prescription
in \cite{Youm:1999zs}, this procedure yielding
\begin{equation}
\label{D0F1Sol}
\begin{array}{lclcccclclc}
H_{\textrm{F}1} & = & \dfrac{Q_{\textrm{F}1}}{r^{6}} & , & & & & H_{\textrm{D}0} & = & 1+Q_{\textrm{D}0}\,
\left(y^{2}+\dfrac{Q_{\textrm{F}1}}{4r^{4}}\right) & ,
\end{array}
\end{equation}
which may be in turn reinterpreted as a background of the class in \eqref{Imamura_Ansatz}, upon performing the following identification
\begin{equation}
\begin{array}{lclcccclclc}
S & = & H_{\textrm{F}1} & , & & \textrm{and} & & K & = & H_{\textrm{D}0} & .
\end{array}
\end{equation}
In the remaining part of this section we will show how to extract massive $\textrm{AdS}_{2}$ solutions from the PDE's \eqref{Massive_Keqn} \& \eqref{Massive_Keqn}.
We will follow the same procedure illustrated in \cite{Bobev:2016phc}, where the $\textrm{AdS}_{7}$ solutions of \cite{Apruzzi:2013yva} are seen as emerging from 
NS5 -- D6 -- D8 systems of the Hanany-Zaffaroni type \cite{Hanany:1997gh}. The key will be understanding which combination of the $(y,r)$ 
coordinates plays the role of the radial coordinate of $\textrm{AdS}_{2}$, thus transforming the original system of PDE's determining $\frac{1}{4}$-BPS
flows into a single ODE yielding warped AdS geometries with enhanced supersymmetry as solutions. 
Analog to the case studied in \cite{Bobev:2016phc}, we will be able to exploit the insight coming from the massless case also when the Romans' mass is turned on.
This will result in an exhaustive classification of all $\textrm{AdS}_2$ of this type in massive type IIA string theory.

\subsection*{The near-horizon limit of the D0 -- F1 system}

Let us first start from the massless solution in \eqref{D0F1Sol} describing the semilocalized intersection between a D-particle and a fundamental string. 
By performing the following coordinate change \cite{Cvetic:2000cj}
\begin{equation}
\left\{
\begin{array}{lclc}
y & = & \dfrac{\cos \theta}{\rho^{2}} & , \\[2mm]
r^{-2} & = & \dfrac{2\sin \theta}{Q_{\textrm{F}1}^{1/2}\rho^{2}} & , 
\end{array}
\right.
\end{equation}
we find that the metric, in the $\rho\rightarrow 0$ limit, takes the form
\begin{equation}
\label{D0F1metric}
ds^2_{10} \,=\, \frac{(Q_{\textrm{D}0} \, Q_{\textrm{F}1})^{1/2}}{8\sin^3\theta} \, \left( \frac{1}{4} ds^2_{{\rm AdS}_2} 
+ d\theta^2 +  4 \sin^2\theta \, d\Omega^2_{(7)}  \right) \ ,
\end{equation}
which is the warped product of $\textrm{AdS}_{2}$ and an 8-manifold obtained as a fibration of $S^{7}$ over a line.

The insight that we can borrow from the massless situation described above is that taking the near-horizon limit where AdS emerges, involves the following
two conditions
\begin{equation}
\begin{array}{lclcclc}
y \ \rightarrow \ 0 & , & r \ \rightarrow \ \infty & , & \textrm{while} & y\,r^{2} \ \sim \ \textrm{finite} & .
\end{array}
\label{NHlimit}
\end{equation}
In the following part of this section we will make use of the above conditions in order to guess the change of variables that translates the PDE's in
\eqref{Massive_Keqn} \& \eqref{Massive_Seqn} into a single ODE to be solved for AdS solutions.

\subsection*{AdS$_2 \times S^7$ solutions in massive type IIA supergravity}

Inspired by \eqref{NHlimit}, we proceed by making the following Ansatz for the $S$ and $K$ functions
\begin{align}
S \,=\, r^{\kappa} \, G(y^{2}r^{4}) \ , \qquad
K \,=\, \frac{2}{ m g_s} \, y\, r^{\kappa+4} \, G'(y^{2}r^{4}) \ ,
\end{align}
where $G$ is an arbitrary function of the combination $y^{2}r^{4}\,\equiv\,\zeta$, and $\kappa$ is a constant yet to be determined. 
For the above choice of $S$ \& $K$ the dilaton reads
\begin{equation}
e^{\Phi} \, = \, \left(\frac{2G'}{m}\zeta^{1/2}\right)^{3/4} \, \left(\frac{g_{s}}{G^{2}}\right)^{1/4} \, r^{\frac{\kappa+6}{4}} \ ,
\end{equation}
which stays finite in the limit \eqref{NHlimit} only when $\kappa=-6$. With this choice for $\kappa$, the 10D metric turns out to be given by 
$\textrm{AdS}_{2}\times S^{7}$ warped over the $\zeta$ coordinate, the warping being specified by the function $G(\zeta)$. 
Upon introducing new coordinates defined by 
\begin{equation}
\left\{
\begin{array}{lclc}
y & = & \rho^{-2} \, f(\zeta) \sqrt{\zeta} & , \\[2mm]
r & = & \rho \, f(\zeta)^{-1/2} & , 
\end{array}
\right.
\end{equation}
where $f^{-4} \equiv G'\, (G + 4 \zeta) \,\sqrt{\zeta}$, one can furthermore check that the full set of equations of motion and Bianchi identities 
are implied whenever $G$ satisfies the following ODE
\begin{equation}
G'' \,+\, \frac{G + 2 \zeta(G'-2)}{2\zeta(G + \zeta)} \, G' = 0 \ ,
\end{equation}
where a prime denotes differentiation with respect to $\zeta$.

This is solved by 
\begin{equation}
\label{ODE_G}
G(\zeta) \,=\, -\frac{3}{8} \gamma_1^2 \,+\, \frac{4}{3} \zeta + \gamma_2 \sqrt{\zeta}\, +
\gamma_2 \sqrt{1 - \frac{16}{3\gamma_1} \sqrt{\zeta}} \ ,
\end{equation}
where $\gamma_1$ \& $\gamma_2$ are integration constants.
The metric now becomes
\begin{equation}
ds^2_{10} \,= \,\sqrt{\frac{2}{m g_s}} (\sqrt{\zeta} G')^{1/2} \left[\frac{G+4\zeta}{\zeta} ds^2_2 \,+ \,\frac{d\zeta^2}{4\zeta(G + 4\zeta)}  + d\Omega_{(7)}^2 \right] \ ,
\end{equation}
with
\begin{equation}
ds^2_2 = - \rho^8 d\tau^2 \,+\, \frac{d\rho^2}{\rho^2}  \ , \qquad \tau \equiv \sqrt{\frac{m g_s}{2}} t \ ,
\end{equation}
where we recognize $ds^2_2$ as the line element of AdS$_2$ of radius $1/4$.

Note that the above solution may be trusted as a good supergravity solution since it admits the $g_s\ll 1$ limit, in which the string dilaton becomes small while
the overall warp factor grows large in string units.

\subsection*{Relation to AdS$_7 \times S^2$ through double analytic continuation}

From the previous analysis it appears evident that our class of $\textrm{AdS}_{2}$ solutions 
realizes the superalgebra $\mathfrak{osp}(8|2)$ as an isometry algebra, just like the $\textrm{AdS}_{7}$ ones in \cite{Apruzzi:2013yva,Apruzzi:2015wna}.
The only difference between the two realizations being the different choice of real form, \emph{i.e.} $\mathfrak{so}(2,1)$ $\oplus$ $\mathfrak{so}(8)$ for AdS$_2$ vs. $\mathfrak{so}(2,6)$ $\oplus$ $\mathfrak{so}(3)$ for AdS$_7$.
A similar phenomenon has been recently discussed in \cite{Corbino:2017tfl} for AdS$_6 \times S^2$ vs. AdS$_2 \times S^6$ solutions of type IIB supergravity, where in both cases 
the aforementioned 8-dimensional geometry is warped over a Riemann surface $\Sigma$.
There it was argued that the two geometries are related by a so-called \emph{double analytic continuation} involving an interchange of AdS and sphere factors,
while at the same time performing a Wick rotation of the coordinate parametrizing the warping.

In order to make a similar relation manifest for our case, we introduce a new coordinate $z$ and a suitable function $\alpha(z)$ such that
\begin{equation}
\zeta = \dot{\alpha}^2 \ , \qquad G = - 4 (\p) \  ,
\end{equation}
where a dot denotes differentiation with respect to $z$. The ODE \eqref{ODE_G} for $G$
then becomes the following one for $\alpha$
\begin{equation}
\ddddot{\alpha} = 0 \ ,
\end{equation}
which is solved cubic polynomials in $z$ of the form
\begin{equation}
\alpha(z) = c_0 + c_1 z + c_2 z^2 + c_3 z^3 \  .
\end{equation}
The metric now becomes
\begin{equation}\label{alphametric}
\ell^{-2} ds^2_{10} \,=\, \left(\frac{\alpha}{\ddot{\alpha}}\right)^{1/2} \left(-\frac{1}{8} \frac{\alpha \ddot{\alpha}}{\p} \,ds^2_{{\rm AdS}_2} \,+\, \frac{\ddot{\alpha}}{8\alpha}\, dz^2 \,+\, d\Omega^2_{(7)}\right) \ ,
\end{equation}
where $ds^2_{{\rm AdS}_2}$ is the line element of AdS$_2$ of unit radius and
\begin{equation}
\ell^4 \,\equiv\, \frac{48 c_3}{m g_s} \,{\rm sgn} (\dot{\alpha})  \  .
\end{equation}
The dilaton reads
\begin{equation}
e^\Phi \,=\, g_s \ell^3 \left( \frac{\alpha}{\ddot{\alpha}}  \right)^{3/4}(-2(\p))^{-1/2} \ .
\end{equation}
We recognize the above metric as the double analytic continuation of the
AdS$_7$ solution of \cite{Apruzzi:2013yva}, \cite{Apruzzi:2015wna} as presented in \cite[Sec. 2.2.3]{Cremonesi:2015bld}.  

Let us note that although the above solution is obtained in massive type IIA supergravity, 
we can get the massless limit by taking $m \to 0$ and $c_3 \to 0$ at the same time. Hence we will discuss the massless solution separately.

\section{Analysis of the solutions}

In this section we analyze the geometry and the dilaton of the solution. To this end, it turns out to be very convenient to use the $z$ coordinate that
directly relates our solutions to the AdS$_7$ ones in \cite{Cremonesi:2015bld}.
We keep $c_3 \neq 0$ as in the opposite case we would obtain a massless solution,
as mentioned in the end of the previous section.

Positivity of the metric \eqref{alphametric} metric requires
\begin{equation}
\alpha \ddot{\alpha} \geq 0 \ , \qquad \p \leq 0 \ .
\end{equation}

The special loci of the geometry, where the warp factor vanishes or 
tends to infinity are (i) $\alpha = 0$, or (ii) $\ddot{\alpha} = 0$, or (iii)
$\p = 0$. Note that, in particular, positivity of the metric requires $\dot{\alpha}$ going to zero, whenever either $\alpha$
or $\ddot{\alpha}$ go to zero. Hence we have the following special loci:
\begin{itemize}
\item a double root of $\alpha$
\item a triple root of $\alpha$
\item a root of $\ddot{\alpha}$ and $\dot{\alpha}$ (stationary point of inflection of $\alpha$)
\item a root of $\p$, while keeping $\ddot\alpha\neq 0$ \& $\dot{\alpha}\neq 0$
\end{itemize}
The discriminant of $\p$ is $\Delta(\p) = -2^8 3^3 c_3^2 \Delta(\alpha)^2$, hence we only need to consider a simple root of $\p$
as a multiple root is also a multiple root of $\alpha$ and this is covered
by the first two cases. In particular $\p$ can have simple roots, a triple root wich
corresponds to a double root of $\alpha$, or a quadruple root which corresponds
to triple root of $\alpha$.

Let us now analyze the geometry and the dilaton near the aforementioned loci.

\begin{itemize}
\item Near a double root $z_0$ of $\alpha$ the metric reads
\begin{equation}
\ell^{-2} ds^2_{10} \,\sim\, \frac{1}{\sqrt{2}} \left( \frac{ \ddot{\alpha}(z_0)}{2^5} ds^2_{{\rm AdS}_2} + d\varrho^2 + \varrho^2 d\Omega^2_{(7)} \right)
\end{equation}
where $\varrho^2 \equiv z-z_0$, and the dilaton stays finite. Hence the internal space becomes $\mathbb{R}^8$ 
and the geometry is regular.

\item Near a triple zero the metric becomes
\begin{equation}
\ell^{-2} ds^2_{10} \,\sim\, \frac{3}{4\sqrt{6}|z-z_0|} \left(  \frac{1}{3} ds^2_{{\rm AdS}_2} + dz^2 + \frac{4}{3}(z-z_0)^2 d\Omega^2_{(7)} \right) \ .
\end{equation}
Then we recognize this as a conical singularity. Its physical interpretation as arising from an orbifold is not clear and hence we will dismiss this case.

\item Near a stationary point of inflection of $\alpha$ the metric reads 
\begin{equation}
\ell^{-2} ds^2_{10} \,\sim\, \sqrt{\frac{c_0}{6}} \left(\frac{3}{4z_0} (z -z_0)^{1/2} dz^2 + (z -z_0)^{-1/2} (ds^2_{{\rm AdS}_2} \,+\ d\Omega^2_{(7)} ) \right) \, .
\end{equation}
The dilaton becomes 
\begin{equation}
\ell^{-3} e^\Phi \,\sim \, \frac{1}{6} \left(\frac{c_0}{6}\right)^{3/4}  (z-z_0)^{-5/4} \ .
\end{equation}
This singularity we recognize as an O8/D8-brane singularity.

\item Near a simple root of $\p$ the metric reads
\begin{equation}
\ell^{-2}ds^2_{10} \,\sim \,\beta_1 \,(z-z_0)^{-1} ds^2_{{\rm AdS_2}} \,+\, \beta_2\, dz^2 \,+\, d\Omega^2_{(7)}  \ .
\end{equation}
The dilaton becomes $\ell^{-3}e^\Phi\,\sim\, \beta_3 (z-z_0)^{-1/2}$.
Here $\beta_1$, $\beta_2$ and $\beta_3$ are constants.
This singularity we recognize as the singularity of a fundamental string smeared over the $S^7$.
\end{itemize}

We now look at the range of $z$. An analysis of the roots of the quartic polymial $\p$ shows that it always has two real roots (simple or multiple). Since the coefficient of the $z^4$ term of $\p$ is $-3 c_3^2$, we conclude that it stays
negative for $z \in [z_0, \infty]$ where $z_0$ is a root. The behavior of the solution at infinity is
\begin{equation}
\label{asymptotic}
\ell^{-2} ds^2_{10} \,\sim\, \frac{1}{\sqrt{6}}\left(\frac{3}{4} \frac{dz^2}{z} \,+\, z\left(\frac{1}{4}ds^2_{{\rm AdS_2}} \,+\,  d\Omega^2_{(7)}\right)\right) \ , 
\end{equation} 
and for the dilaton $\ell^{-3} e^\Phi \,\sim \,\beta_4 \,z^{-1/2}$, with $\beta_4$ a constant.
 
It is worth mentioning that such a behavior at infinity can be understood as that of a curved D8-domain wall carrying D0-brane charge. 
This type of BI-on solutions was also investigated in \cite{Imamura:2001cr}. To make our claim manifest, we describe our asymptotic solution at $\infty$
as
\begin{align}
S_{\infty} \,\equiv\, \frac{4}{3}\,\frac{y^{2}}{r^{2}} \ , \qquad
K_{\infty} \,\equiv\, \frac{8}{ m g_s} \, \frac{y}{r^{2}} \ ,
\end{align}
which corresponds to picking $G_{\infty} \,=\, \frac{4}{3} \zeta$, and interpret it as a curved domain wall solution whose profile is
described by
\begin{equation}
S^{-1/2}\,dy \, = \, \frac{1}{\sqrt{3}} \, dr \ ,
\end{equation}
which for $S\,=\,S_{\infty}$ integrates to $y\,=\,c\,r^{2/3}$. One can then check that the metric \eqref{asymptotic} is reproduced by the following
change of coordinates
\begin{equation}
\left\{
\begin{array}{lclc}
y & = & 4z^{1/2} \, e^{-2R} & , \\[2mm]
r & = & z^{3/4} \, e^{R} & , 
\end{array}
\right.
\end{equation}
where $R$ is to be identified with the $\textrm{AdS}_{2}$ radial coordinate. 

\subsubsection*{Compactifying the range of the warp coordinate}
Although $z$ is defined on a half-line we can compactify it by making a coordinate tranformation to a trigonometric function. Let us look at the case where $\alpha$ has
a double root, in the neighborhood of which the internal space is regular. We thus take
\begin{equation}
 \alpha \,=\, c_3 (z-z_0)^2 (z-z_1) \nonumber
\end{equation}
and perform the following coordinate transformation
\begin{equation}
z \,\equiv\, (z_0 - z_1) \tan^2\theta + z_0 \ , \quad \textrm{with}  \qquad \theta \in \left[0, \frac{\pi}{2}\right] \ .
\end{equation}
The metric then becomes
\begin{equation}
\ell^{-2} ds^2_{10} \,= \, \frac{|z_0-z_1|}{\sqrt{2}} \frac{\sec^2\theta}{\sqrt{1+2\sin^2\theta}} \left(\frac14\frac{1 + 2\sin^2\theta}{4-\sin^2\theta} ds^2_{{\rm AdS}_2} + \frac{1+2\sin^2\theta}{\cos^2\theta} d\theta^2 + \sin^2\theta d\Omega^2_{(7)}\right) \ .
\end{equation}

\subsubsection*{The massless solution}
In the massless limit we need to take $c_3 = 0$ and so $\alpha$ becomes a quadratic polynomial. $\p$ evaluates to $c_1^2 - 4 c_0 c_2$ which is also the discriminant of 
$\Delta(\alpha)$ of $\alpha$. Since $\p$ has to be negative we conclude that 
$\alpha$ has complex roots and is always different from zero.
The metric now reads
\begin{equation}
ds^2_{10} \,=\, \left( \frac{\alpha}{2 c_2} \right)^{1/2} \left(- \frac{c_2}{4} \frac{\alpha}{\Delta(\alpha)} ds^2_{{\rm AdS}_2}  + \frac{c_2}{4} \frac{1}{\alpha} dz^2 + d\Omega^2_{(7)}\right) \ .
\end{equation}

Upon making the following coordinate transformation
\begin{equation}
\sin^2\theta \,\equiv\, -\frac{\Delta(\alpha)}{4 c_2} \frac{1}{\alpha} \ ,
\end{equation}
The metric takes exactly the form in \eqref{D0F1metric} obtained as the near-horizon geometry of the massless D0 -- F1 semilocalized intersection in \cite[Sec. 4.5]{Cvetic:2000cj}, 
provided that $Q_{\textrm{D}0} \equiv \frac{1}{2 c_2^2}$, $Q_{\textrm{F}1} \equiv -\Delta(\alpha)$.

\section{Conclusions}

In this paper we have studied a class of warped supersymmetric $\textrm{AdS}_{2}$ solutions in massive type IIA emerging from D0 -- F1 -- D8 systems as 
their near-horizon geometries. All solutions in this class turn out to preserve sixteen real supercharges and the corresponding internal geometry is given by a 
fibration of a (round) seven-sphere over a line. 

The solutions can be on the one hand obtained as special cases of the $\frac14$-BPS family of backgrounds studied in \cite{Imamura:2001cr}, and on the other hand 
reinterpreted as the double-analytically continued version of the $\textrm{AdS}_{7}$ solutions of \cite{Apruzzi:2013yva,Apruzzi:2015wna}. 
This latter relation is further corroborated by the fact that both families offer explicit realizations of the $\mathfrak{osp}(8|2)$ superalgebra within
massive type IIA supergravity, their distinction just corresponding to different choices of real form. 

Due to the corresponding brane picture of the vacua discussed here, we expect them to be dual to some $\mathcal{N}=(8,0)$ superconformal quantum mechanics.
The first interesting check for a possible $\textrm{AdS}_{2}/\textrm{CFT}_{1}$ correspondence in this context would be to be able to compute the holographic
central charge by following the prescription
\begin{equation}
c_{\textrm{hol}} \, \propto \, \frac{L_{\textrm{AdS}}^{d-1}}{G_{\rm N}^{(d+1)}} \ ,
\nonumber
\end{equation}
where both the effective AdS radius $L_{\textrm{AdS}}$ and the Newton constant $G_{\rm N}^{(d+1)}$ are averaged over the warping, and make sense of the 
answer from a field theory viewpoint.

However, it may be worth mentioning that the above prescription for computing $c_{\textrm{hol}}$ would yield an \emph{infinite result} in our case. 
This is mainly due to the range of the $z$ warp coordinate being non-compact. 
Similar pathologies where also encountered in \cite{Lozano:2016kum,Itsios:2017cew} in the context of $\textrm{AdS}_{5}$ solutions in type IIA obtained by
taking a non-Abelian T-duality (NATD) on the type IIB Klebanov-Witten $\textrm{AdS}_{5}\times T^{1,1}$ background \cite{Klebanov:1998hh}.
There this issue was resolved on the gravity side by introducing a hard cut-off which should be interpreted from a physical perspective as inserting
the suitable branes which interrupt the dual quiver which would otherwise be infinitely long. 

It would be interesting to investigate whether this feature that our solutions have, can be resolved
in a way similar to the NATD solutions. 
This could possibly shed a light on their holographic dual superconformal quantum mechanical models and the possible emergence of deconstructed extra 
dimensions underlying this structure. We hope to come back to these issues in the future.

\section*{Acknowledgements}

We would like to thank Y.~Lozano, N.~Petri, L.~Tizzano and A.~Tomasiello for very stimulating discussions, and C.~Uhlemann for useful correspondence.
The work of GD is supported by the Swedish Research Council (VR). The work of AP is supported by the Knut and Alice Wallenberg Foundation under grant Dnr KAW 2015.0083.

\bibliographystyle{utphys}
\bibliography{AdStwo}

\end{document}